\documentclass{ws-procs9x6}
\begin{document}
\title{Deeply Virtual Compton Scattering with {\large\tt CLAS12}.}
\author{ L. Elouadrhiri$^*$}
\address{Jefferson Lab\\
Newport News, VA 23606, USA\\
$^*$E-mail: latifa@jlab.org}

\begin{abstract}
An overview is given about the capabilities provided by the JLab 12 GeV Upgrade to
measure deeply virtual exclusive processes with high statistics and covering a large 
kinematics range in the parameters that are needed to allow reconstruction of a spatial 
image of the nucleon's quark structure. The measurements planned with {\tt CLAS12} will
cross section asymmetries with polarized beams and with longitudinally and transversely 
polarized proton targets in the constrained kinematics $x = \pm \xi$. 
In addition, unpolarized DVCS cross sections, and doubly polarized 
beam target asymmetries will be measured as well. In this talk only the beam and target 
asymmetries will be discussed.          
\end{abstract}
\keywords{GPDs, DVCS}
\bodymatter
\section{Introduction}

The concept of GPDs~\cite{mueller:1994,Ji:1996nm,Radyushkin:1996nd} has led to completely 
new methods of ``spatial imaging''
of the nucleon.
The mapping of the nucleon GPDs, and a detailed understanding of the spatial quark 
distribution of the nucleon, have been widely 
recognized are a key objectives of nuclear physics of the 
next decade, and is a key justification for the JLab energy upgrade to 12 GeV. 
GPDs also allow to quantify how the orbital motion of quarks in the nucleon 
contributes to the nucleon spin -- a question of crucial importance for our 
understanding of the ``mechanics'' underlying nucleon structure. 
This requires a comprehensive program, combining results
of measurements of a variety of processes in electron--nucleon 
scattering with structural information obtained from theoretical studies, as 
well as with expected results from future lattice QCD simulations.

\begin{figure}[htb]
\resizebox{0.9\textwidth}{!}{%
  \includegraphics{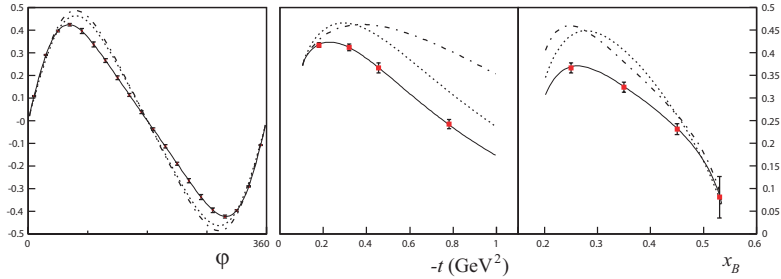}}
\caption{The beam spin asymmetry showing the DVCS-BH interference for 11 GeV beam 
energy from the VGG model~\cite{vgg} with uncertainties projected for 11 GeV~\cite{e12-06-119}. 
Many other bins will be measured simultaneously.}
\label{fig:dvcs_alu_12gev}   
\end{figure}

It is well recognized~\cite{Belitsky:2001ns,Burkardt:2002hr,Belitsky2004} that 
exclusive processes can be used to probe the GPDs and construct 2-dimensional and 
3-dimensional images of the quark content of the nucleon. Deeply virtual Compton 
scattering and deeply virtual meson production are identified as the processes most suitable to 
map out the twist-2 vector GPDs $H,~E$ and the axial GPDs ${\tilde H},~{\tilde E}$ in $x,~\xi,~t$, 
where $x$ is the momentum fraction of the struck quark, $\xi$ the longitudinal momentum 
transfer to the quark, and $t$ the momentum transfer to the nucleon. Having 
access to a 3-dimensional image of the nucleon (two dimensions in transverse space, one 
dimension in longitudinal 
momentum) opens up completely new insights into the complex structure of the nucleon. 
For example, the nucleon matrix element of the energy-momentum tensor contains 3 form 
factors that encode information on the angular momentum distribution $J^q(t)$ of the 
quarks with flavor $q$ in transverse space, their mass-energy distribution $M_2^q(t)$, 
and their pressure and force distribution $d^q_1(t)$. These form factors also appear as 
moments of the vector GPDs~\cite{goeke2007}, thus offering prospects of accessing these 
quantities through detailed mapping of GPDs. The 
quark angular momentum in the nucleon is given by 
$$J^q(t) = 
\int_{-1}^{+1}dx x [H^q(x, \xi, t) + E^q(x, \xi, t)]~,$$ and the mass-energy and pressure distribution $$M_2^q(t) + 4/5d^q_1(t)\xi^2 
= \int_{-1}^{+1}dx x H^q(x, \xi, t)~.$$ The mass-energy and force-pressure distribution 
of the quarks are given by the second moment of GPD $\it{H}$, and their relative contribution is controlled by $\xi$. The separation of $M^q_2(t)$ and 
$d^q_1(t)$ requires measurement of these moments in a large range of 
$\xi$.

\begin{figure}[t]
\begin{center}
\psfig{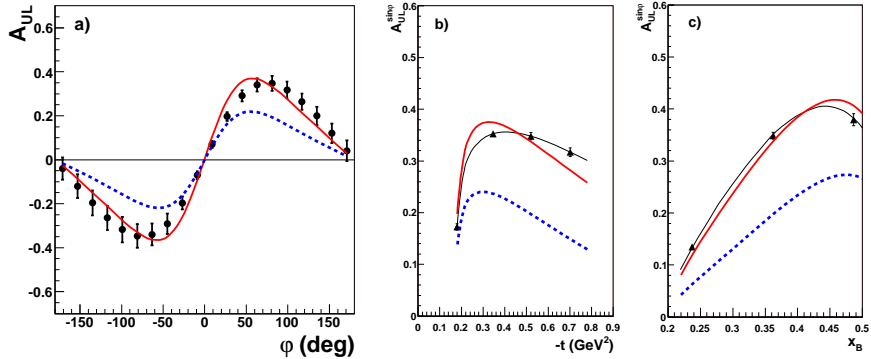}
\end{center}
\caption{The longitudinal target spin asymmetry beam spin asymmetry showing the DVCS-BH 
interference for 11 GeV beam energy from the VGG model~\cite{vgg} with uncertainties 
projected for 11 GeV~\cite{e12-06-119}. Other bins will be measured simultaneously }
\label{fig:aul}
\end{figure}

\section{GPDs and DVCS}

DVCS has been shown~\cite{clas-dvcs-1,clas-dvcs-2,halla-dvcs,clas-dvcs-3} to be the 
cleanest process to access GPDs at the kinematics accessible today. It is also 
a relatively rare process and requires high luminosities for the required high 
statistics measurements. The beam helicity-dependent cross section asymmetry is 
given in leading twist as 
$$ A_{LU} \approx \sin\phi[F_1(t)H + \xi(F_1(t)+F_2(t))\tilde{H}]d\phi~, 
$$where $F_1$ and $F_2$ are the 
Dirac and Pauli form factors, $\phi$ is the azimuthal angle between the electron scattering plane 
and the hadronic plane. The kinematically suppressed term with GPD $E$ is omitted. 
For not too large $\xi$ the asymmetry is mostly sensitive to the GPD $H(x=\xi,\xi,t)$.

The asymmmetry with a longitudinally polarized target is given by  
$$A_{UL} \approx \sin\phi[F_1(t)\tilde{H} + \xi(F_1(t)+F_2(t)H]~.$$  
The combination of $A_{LU}$ and $A_{UL}$ allows a separation of GPD $H(x=\xi,\xi,t)$ and
$\tilde{H}(x=\xi,\xi,t)$.  

Using a transversely polarized target the asymmetry 
$$A_{UT} \approx \sin\phi t/4M^2 [F_2(t)H - F_1(t)E] $$ can be measured, which 
depends in leading order on GPD $E$ and is highly sensitive to orbital angular 
momentum contributions of quarks.  

Clearly, determining moments of GPDs for different $t$ will require measurement 
in a large range of $x$, in particular at large $x$. 
The reconstruction  of the transverse spatial quark distribution requires measurement in 
a large range in $t$, and the separation of the $d^q_1(t)$ and $M^q_2(t)$ form factors 
requires a large span in $\xi$.    

\section{Upgrade of CLAS to {\large\tt CLAS12}.}

To meet the requirements of high statistics measurements of relatively rare 
exclusive processes such as DVCS at high photon virtuality $Q^2$, large $t$ and $\xi$, 
the CLAS detector will be 
upgraded and modified to {\tt CLAS12}~\cite{clas12-tdr}. The main new features of {\tt CLAS12} over the current 
CLAS detector include a high operational luminosity of $10^{35}$cm$^{-2}$sec$^{-1}$, 
an order of magnitude increase over CLAS~\cite{clas}. Improved particle identification 
will be achieved with additional threshold gas Cerenkov counter, improved timing resolution 
of the forward time-of-flight system, and a finer granularity electromagnetic preshower 
calorimeter that, in conjunction with th existing CLAS calorimeter will provide much 
improved $\gamma/\pi^0$ separation for momenta up to 10 GeV. In addition, a new central 
detector will be built 
that uses a high-field solenoid magnet for particle tracking and allows the operation of
dynamically polarized solid state targets. With these upgrades {\tt CLAS12} will be 
the workhorse for exclusive electroproduction experiments in the deep inelastic kinematics.

\begin{figure}
\resizebox{0.8\textwidth}{!}{%
  \includegraphics{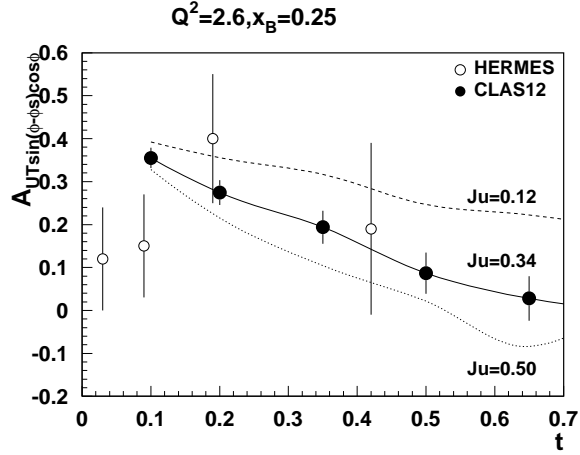}}
\caption{Projected transverse target asymmetry $A_{UT}$ for DVCS production off protons at 11 GeV beam energy.
The curves represent different assumptions on the u-quark contributions to $J(t)$.}
\label{fig:dvcs_aut_12gev}    
\end{figure}

\section{Projected results at 12 GeV with {\tt\large CLAS12}.}

The 12 GeV upgrade offers much improved capabilities to access GPDs. 
Figure~\ref{fig:dvcs_alu_12gev} shows the expected statistical precision of 
the beam DVCS asymmetry for some sample kinematics. At the expected luminosity 
of $10^{35}$cm$^{-2}$sec$^{-1}$ and for a run time of 2000 hours, 
high statistics measurements in a very large kinematics range are possible. 

Using a dynamically polarized $NH_3$ target we can also measure the longitudinal target 
spin asymmetry $A_{UL}$ with high precision. The projected results are shown in 
Fig.~\ref{fig:aul}. The statistical accuracy of this measurement will be less 
than for the $A_{LU}$ asymmetry due to the large dilution factor in the
target material, but it will still be a very significant measurement.  

Polarizing the target transverse to the beam direction will access a different combination
of GPDs, and provide different sensitivity for the y- and x-components of the target 
polarization. The expected accuracy for one of the polarization projections is shown in 
Fig.~\ref{fig:dvcs_aut_12gev}. Here the target is assumed to be a frozen HD-Ice target, 
which has different characteristics from the $NH_3$ target.     

\begin{figure}
\resizebox{0.7\textwidth}{!}{%
  \includegraphics{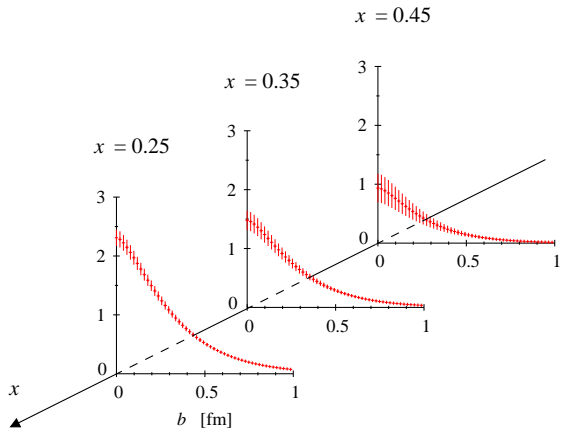}}
\caption{The u-quark distribution in transverse space as extracted from projected DVCS data with {\tt CLAS12}. }
\label{fig:gpd_H}    
\end{figure}

A measurement of all 3 asymmetries will allow a separate determination of GPDs 
$H,~\tilde{H}$ and $E$ at the above specified kinematics. 
Through a Fourier transformation the t-dependence of GPD $H$ can 
be used to determine the $u-$quark distribution in transverse impact parameter space. 
Figure~\ref{fig:gpd_H} shows projected results for such a transformation assuming a model
parameterization for the kinematical dependences of GPD $H$. Knowledge of GPD $E$ will be 
particularly interesting as it is directly related to the orbital angular momentum distribution
of quarks in transverse space.

\vspace{0.5cm}
\noindent{\bf Acknowledgment}

We thank the members of the CLAS collaboration who contributed to the 
development of the exciting physics program for the JLab upgrade to 12 GeV, 
and the {\tt CLAS12} detector. Much of the material in this report is taken 
from the {\tt CLAS12} Technical Design Report Version 3, October 2007~\cite{clas12-tdr}.  

This work was supported in part by the U.S. Department of Energy and the National Science Foundation, the French Commisariat {\'{a}} l'Energie Atomique, the Italian Instituto Nazionale di Fisica Nucleare, the Korea Research Foundation,  and a research grant of the Russian Federation. The Jefferson Science Associates, LLC, operates Jefferson Lab under contract DE-AC05-060R23177.


\begin{thebibliography}{99}
\bibitem{clas12-tdr} The complete Technical Design Report document may be obtained from the authors.
\end{thebibliography}
\end{document}